\documentclass[preprint2,trackchanges]{aastex6}
\usepackage{epsfig}
\usepackage{gensymb}
\usepackage{amsmath}
\usepackage{ulem}
\usepackage{verbatim}
\usepackage{commath}
\begin{document}
\newcommand{\cd}{CO$_2$}
\newcommand{\cm}{cm$^{-1}$}
\title{Diffusion and Clustering of Carbon Dioxide on non-porous Amorphous Solid Water}

\author{Jiao He\altaffilmark{1} and Shahnewaj M. Emtiaz and Gianfranco Vidali\altaffilmark{2}}
\affil{Physics Department, Syracuse University, Syracuse, NY 13244, USA}
\altaffiltext{1}{jhe08@syr.edu}
\altaffiltext{2}{gvidali@syr.edu}
\begin{abstract}
Observations by ISO and Spitzer towards young stellar objects (YSOs) showed that CO$_2$ segregates in the icy mantles covering dust grains. Thermal processing of ice mixture was proposed as responsible for the segregation. Although several laboratory studied thermally induced segregation, a satisfying quantification is still missing. We propose that the diffusion of CO$_2$  along pores inside water ice is the key to quantify segregation. We combined Temperature Programmed Desorption (TPD) and Reflection Absorption InfraRed Spectroscopy (RAIRS) to study how CO$_2$ molecules interact on  a non-porous amorphous solid water (np-ASW) surface. We found that CO$_2$ diffuses significantly on a np-ASW surface above 65~K and clusters are formed at well below one monolayer. A simple rate equation simulation finds that the diffusion energy barrier of CO$_2$ on np-ASW is 2150$\pm$50 K, assuming a diffusion pre-exponential factor of 10$^{12}$ s$^{-1}$. This energy should also apply to the diffusion of CO$_2$ on wall of pores. The binding energy of CO$_2$ from CO$_2$ clusters and CO$_2$ from H$_2$O ice have been found to be $2415\pm20$ and $2250\pm20$~K, respectively, assuming the same prefactor for desorption. CO$_2$-CO$_2$ interaction is stronger than CO$_2$-H$_2$O interaction, in agreement with the experimental finding that CO$_2$ does not wet np-ASW surface. For comparison, we carried out similar experiments with CO on np-ASW, and found that the CO-CO interaction is always weaker than CO-H$_2$O. As a result, CO wets np-ASW surface. This study should be of help to uncover the thermal history of CO$_2$ on the icy mantles of dust grains.
\end{abstract}

\keywords{astrochemistry --- ISM: molecules --- methods: laboratory: solid state --- methods: laboratory: molecular --- dense matter}

\section{Introduction}\label{sec:intro}
\cd{} constitutes a significant fraction of ices coating dust grains in interstellar clouds, with \cd{} abundance ranging between 15\% and 40\%.  Other major ice components include water, the most abundant, and CO and CH$_3$OH \citep{Pontoppidan2008,Boogert2015}. Through observations of solid state of CO$_2$ and CO IR features  a picture of the state and history of the cloud \citep{Nummelin2001,Whittet1998,Whittet2007,Whittet2009,Boogert2015,Cook2011} can be obtained.
By analyzing the double-peak feature of \cd{} bending mode at 660 cm$^{-1}$, \cd{} in the pure form has been found by ISO \citep{Gerakines1999} and Spitzer \citep{Pontoppidan2008} towards young stellar objects (YSOs). The fact that this double-peak feature is only observed toward regions where the protostar has already ``turned on'' (see \citet{Oberg2009,Pontoppidan2008,Boogert2015} and references therein) suggests that the segregation of \cd{} is induced by thermal processing of the ice mixture.
To explain these observations and characterize the segregation of \cd{} in ices,  the following three mechanisms are possible: (a) collapse of the porous ice structure that causes  \cd{} molecules to move and get trapped in pockets of high \cd{} concentration; (b) penetration of \cd{} molecules through bulk non-porous amorphous solid water ice (np-ASW) by diffusion; (c) diffusion of \cd{} along the wall of pore surfaces. The first one requires highly porous structure and relatively high concentration of \cd{} with respect to water, a scenario not yet confirmed by observations \citep{Keane2001}. The second one requires a low energy barrier for diffusion in the bulk, but experiments show otherwise \citep{Oberg2009,Mispelaer2013}. The third one --- diffusion along the surfaces of the walls of ice --- is the most probable mechanism for \cd{} segregation, and is the focus of this study.

There have been several laboratory works that studied mid-IR features of \cd{}, i.e., the bending mode $\nu_2$ at 660 cm$^{-1}$ and the asymmetric stretch band $\nu_3$ at $\sim$ 2340 cm$^{-1}$ and  near-IR overtones and combination bands \citep{Ehrenfreund1998,Dartois1999,Baratta2000,Palumbo2000,Bernstein2005,Hodyss2008,Oberg2009,Isokoski2014}.
Studies of \cd{}:H$_2$O binary and \cd{}:H$_2$O:CH$_3$OH ternary ices have found that segregation becomes significant above 50--60~K and up to 85~K \citep{Ehrenfreund1998,Palumbo2000,Bernstein2005,Hodyss2008}, although quantitative values so obtained differ among these measurements (these temperature values refer to laboratory experimental conditions; in space they would be lower, see below). \citet{Palumbo2000}  studied  the dependence of the mixing ratio \cd{}:H$_2$O on segregation, and found that below 10\% \cd{} in H$_2$O there is no segregation of \cd{}. \citet{Oberg2009} did a comprehensive study of segregation in binary mixtures (\cd{}:H$_2$O); they found a thickness as well as a composition dependence. Coupling data with a Monte Carlo simulation, they reported an energy barrier  of 1080 $\pm$ 190~K for segregation of \cd{} at the surface of the ice. \citet{Isokoski2014} studied segregation of \cd{} in H$_2$O using the combination bands $\nu_1+\nu_3$ at 3704 cm$^{-1}$ and $2\nu_2+\nu_3$ at 3597 cm$^{-1}$ which are sensitive to the local environment \citep{Keane2001}. Their study, which explored the influence of morphology of the ice, showed that high porosity facilitates \cd{} segregation. Indeed this and other studies suggest the link between diffusion of \cd{} molecules on the surface of ice and segregation. The work presented here addresses this issue in a quantitative way.

Surface diffusion of \cd{} on ice (whether on pores or at the interface between the ice surface and vacuum) is much faster than {\em through} ice \citep{Oberg2009}; thus, recent studies were concentrated on studying diffusion and aggregation on the surface of water ice.
\citet{Ghesquiere2015} studied in the laboratory the diffusion of \cd{} through compact ASW, but in the experiments the ice was deposited at 80~K, which may not be high enough to be truly compact. On the contrary, our laboratory measurements show no evidence of \cd{} penetration into np-ASW deposited at 130~K. Another important factor for the formation of \cd{} clusters is the binding energy of \cd{} from \cd{} clusters and \cd{} from water ice surface. Clusters can be formed only when the binding energy of \cd{} from \cd{} clusters is stronger than that of \cd{} from water ice surface. To study the binding energies,  \citet{Noble2012} used temperature programmed desorption (TPD) to measure the TPD of CO$_2$ on  a np-ASW  surface down to 0.5 monolayer (ML) coverage; based on the argument that \cd{} interaction energy with the water ice surface is stronger than the interaction energy at monolayer coverage,  they argued for non-wetting. However, the measurement at 0.5 ML coverage  is not low enough to study the onset of segregation or cluster formation (see below); therefore the behavior of \cd{} at lower coverages is unknown.  \citet{Karssemeijer2014} used molecular dynamics simulation with a new ab initio calculation of the interaction energy of CO$_2$-H$_2$O and \cd{}-\cd{} in the gas phase and on water ice. They found that in the gas phase the CO$_2$-H$_2$O interaction strength is twice as  the \cd{}-\cd{}, while in the solid state the difference between them is much smaller. The result of their molecular dynamics simulations found no indication of island formation or clustering during deposition of \cd{} on water ice. Because the simulation occurs on time scales orders of magnitude faster than an actual deposition, some caution should be exercised in applying these results.

To resolve this issue and obtain information of how \cd{} molecules diffuse on water ice, as well as the binding energies, we used a combination of TPD and RAIRS techniques to  study systematically the interaction of \cd{} with np-ASW\@. For comparison purposes, some of the experiments were also done with CO\@. We explored both the temperature and coverage dependence of \cd{} clustering on np-ASW, and found how the diffusion rate determines the clustering of \cd{}. An estimate of the energy barrier to diffusion of \cd{} molecules on water ice is also obtained. Technical improvements allowed us to study  the \cd{}-H$_2$O system down to $\sim1$\% ML coverage of \cd{}. This is an important step that allows us to better approximate the conditions of the experiment to the conditions in the ISM\@. By choosing a np-ASW ice, we single out the interaction between \cd{} and the surface of water, and we avoid the complication of the structural changes of porous ice and of the trapping of \cd{} in pores that affect the segregation rate. However, our results should be translatable to  more porous ice surfaces.

The remainder of the paper is organized as below: in Section~\ref{sec:exp} we describe briefly the experimental setup and how ices are grown and characterized; Section~\ref{sec:result} presents the results and analysis, followed by a discussion of the astrophysical implications in Section~\ref{sec:dastro}.

\section{Experimental Setup}\label{sec:exp}
Experiments were carried out in an ultra-high vacuum (UHV) chamber connected to two molecular beam lines. The pressure in the UHV chamber reaches $1\times10^{-10}$ torr after bake-out. Ices are grown on a gold coated copper disk located at the center of the UHV chamber. The sample disk can be cooled by liquid helium to $\sim 8$~K and heated up to 450~K by a resistive heater. The temperature of the sample disk was measured by a calibrated silicon diode (Lakeshore DT 670) and controlled by a Lakeshore 336 temperature controller with an uncertainty of less than 50~mK. A Hiden Analytical quadrupole mass spectrometer (QMS) was mounted on a rotatable flange that can be used to measure the intensity of the incoming molecular beam or the molecules desorbing from the sample in Temperature Programmed Desorption (TPD) experiments. A Nicolet 6700 Fourier Transform Infrared (FTIR) spectrometer in Reflection Absorption InfraRed Spectroscopy (RAIRS) setup was used to monitor ices grown on the substrate. The infrared beam was focused by an off-axis paraboloidal mirror before entering the chamber through a differentially pumped KBr window. It is then reflected from the sample at $78\degree$ incidence angle and exits the chamber through another KBr window,; it is then focused by an off-axis ellipsoidal mirror before entering a MCT-B detector with a working range from 650 to 4000~cm$^{-1}$. For experiments carried out in this work, the highest resolution 1~cm$^{-1}$ was used. Spectra were collected by averaging 22--23 scans every 30 seconds.

Water ice was grown by vapor deposition from the background using a capillary array \citep{He2016b}. The capillary array is facing empty space in the chamber for background deposition onto the substrate. A nude ionization pressure gauge was used to monitor the pressure in the chamber. We assume that during background deposition, the pressure readout from the pressure gauge is the same as that close to the substrate.
Distilled water was used for water deposition. It went through at least three freeze-pump-thaw cycles before being introduced into the chamber.
In this work we mostly use non-porous amorphous solid water (np-ASW), which was grown by background deposition of water vapor when the substrate was at 130 K. After deposition, the ice remained at 130 K for 20 minutes to stabilize the ice structure before cooling down for further experiments.
CO$_2$ ice was grown either by deposition from the molecular beam or by background deposition from the capillary array. The thickness of CO$_2$ ice grown from background deposition was calculated by integrating the chamber pressure over time. The ion gauge correction for different gases was taken into account. The recording of the pressure in the chamber  and the calculation of the thickness (in Langmuir) in real time were handled by a LabVIEW program. This ensures that the thickness of deposition can be controlled accurately within an uncertainty of 3\%.  Unity sticking is assumed for all of the experiments in this work, and this is supported by previous sticking measurement by our group \citep{He2016a}. At room temperature ($\sim$294~K), 1 Langmuir (1L= $1\times10^{-6}$ torr$\cdot$s) of H$_2$O, CO, and CO$_2$ amounts to $5.23\times10^{14}$, $4.19\times10^{14}$, and $3.34\times10^{14}$  molecules cm$^{-2}$, respectively --- see the Appendix for details. The column density of \cd{} in  units of L can be converted to thickness as 1~L=1.9\AA, assuming a \cd{} density of 1.28 g$\cdot$cm$^{-3}$ \citep{Gerakines2015}. From CO TPD spectra in Section~\ref{sec:result} we  show that 1~L of CO corresponds to about 1 monolayer (ML) coverage on np-ASW\@. We assume that the number of \cd{} molecules in 1~ML coverage is the same as that of CO, and in the remaining part of this work we use L and ML interchangeably for both \cd{} and CO\@.

Details of the molecular beam can be found in a previous publication \citep{He2016a}. The deposition rate from the beam was controlled by an Alicat MCS-5 mass flow controller with a relative uncertainty of less than 1\%. The beam deposition time was automated using a computer controlled flag with an accuracy down to $\sim$50 ms. The molecular beam is at $8\degree$ with respect to surface normal, and covers most of the sample surface area, which is 1 cm$^2$. Because the infrared beam does not overlap perfectly with the molecular beam on the sample, the calibration of thickness needs special attention. Usually the ice thickness can be calculated by integrating the absorption of a known absorption band and comparing it with the absorption intensity found in the literature. But this method does not apply here, because the molecular beam covers a smaller area than the infrared beam.
We instead compared the infrared spectrum of CO$_2$ from beam deposition and from background deposition on annealed water ice to calibrate the beam deposition rate/beam flux. More details of the calibration is in the Appendix.

\section{Results and Analysis}\label{sec:result}
\begin{figure}
    \epsscale{1.1}
\plotone{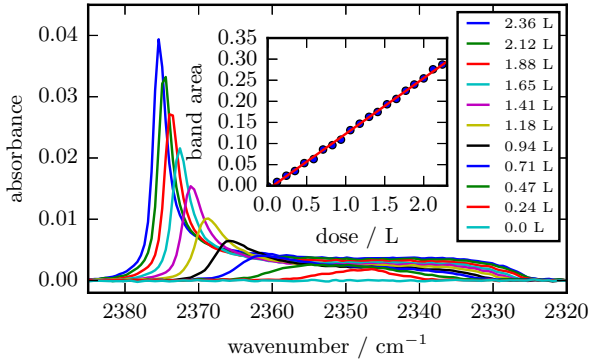}
\caption{CO$_2$ $\nu_3$ band during deposition of \cd{} on np-ASW at 10 K. The integrated band area versus deposition dose is shown in the center inset. The deposition dose for each curve is in the right side inset.\label{fig:co2_com_dep}}
\end{figure}

In the first experiment, \cd{} was deposited from the beam onto the np-ASW ice at 10~K. Infrared spectra were recorded during deposition, see Figure~\ref{fig:co2_com_dep}. At very low dose, \cd{} molecules are isolated on the surface and interact primarily with the water surface, as indicated by the asymmetrical stretch $\nu_3$ absorption peak at 2347 \cm{}. As the  CO$_2$ concentration increases, CO$_2$ molecules interact primarily among themselves in clusters. In the RAIRS geometry of absorption and reflection at glancing angle, solid state \cd{} aggregates are characterized by a peak at $\sim$ 2380 cm$^{-1}$ \citep{Escribano2013,Edridge2013} due to the LO phonon mode \citep{Berreman1963}. Also shown in Figure~\ref{fig:co2_com_dep} is the integrated area of $\nu_3$ band, which is linear with deposition dose, regardless of the peak shape and position.

\begin{figure}
    \epsscale{1.1}
\plotone{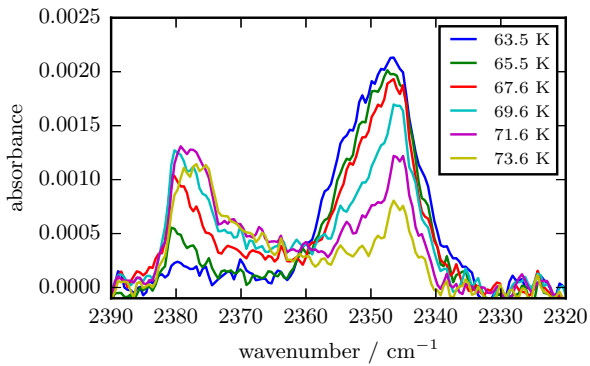}
\caption{CO$_2$ $\nu_3$ band during heating in the following experiment: 0.23 L of CO$_2$ was deposited from the beam onto np-ASW at 10 K, followed by heating  at 2~K/minute. Surface temperature is shown in the inset.\label{fig:co2_com_6}}
\end{figure}

In the second experiment, 0.23 L of \cd{} was deposited from the beam at a rate of 0.35 L$\cdot$minute$^{-1}$ on np-ASW surface at 10~K, and it was followed by a linear increase of the temperature at 2~K$\cdot$minute$^{-1}$. Figure~\ref{fig:co2_com_6} shows the spectra of $\nu_3$ band during heating. Compared with \citet{Oberg2009}'s and \citet{Noble2012}'s work, we focus on a much lower surface coverage of \cd{} in order to find the temperature at which \cd{} starts to diffuse on the surface. At low temperature and low surface coverage, the isolated \cd{} molecules have an absorption band at around 2347 \cm{}. As the sample is heated  to $\sim$65~K, the band at 2347 \cm{} decreases and another band emerges at $\sim$2380 \cm{}, which is associated with aggregated \cd{}. This indicates that at $\sim$65~K, the diffusion rate of \cd{} on np-ASW becomes significant and CO$_2$ molecules are able to move around and form clusters. The IR spectrum therefore shows a decrease in the band of \cd{}-water interaction  and an increase in the one of  \cd{}-\cd{} interaction. From this experiment we found that in the laboratory time scale, the ice temperature needs to be  close to or higher than 65~K for \cd{} segregation/clustering to happen.

\begin{figure}
    \epsscale{1.1}
\plotone{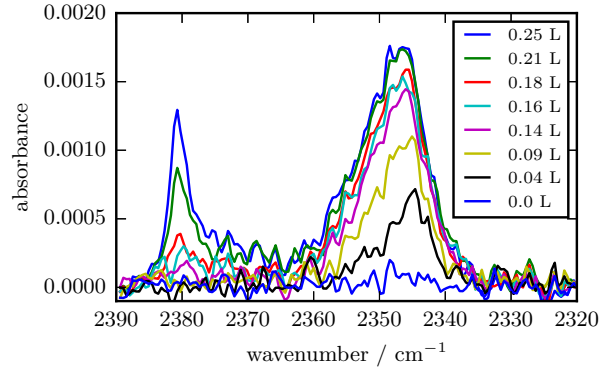}
\caption{CO$_2$ $\nu_3$ band during deposition of CO$_2$ from the beam onto a np-ASW at 65 K. The deposition dose is shown in the inset.\label{fig:co2_com_65k_dep}}
\end{figure}

Clustering behavior of \cd{} also depends on surface coverage. The more \cd{} is on the surface, the easier is to form clusters. At a given temperature, there exists a threshold coverage above which \cd{} forms clusters. We carried out the following experiment to find out the clustering threshold at 65~K. With np-ASW at 65 K we deposited \cd{} continuously at a rate of 0.035 L$\cdot$minute$^{-1}$ and monitored the IR spectrum. At this surface temperature \cd{} should be mostly mobile. Figure~\ref{fig:co2_com_65k_dep} shows that there is only one peak at 2347 \cm{} below $\sim$0.17~L, while at and above $\sim$0.17~L the peak at 2380~\cm{} begins to emerge. Therefore 0.17~L is the clustering threshold at 65~K. The existence of a coverage threshold for clustering is consistent with \citet{Palumbo2000}, which found a concentration threshold of 10\% for clustering. At higher surface temperature, the threshold can be lower.

\begin{figure}
\plotone{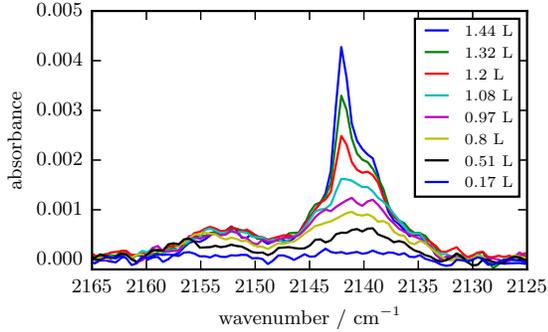}
\caption{CO stretching band during deposition of CO from the beam on np-ASW at 22~K. The deposition dose is shown in the inset.\label{fig:co_com_spectrum}}
\end{figure}

\begin{figure}
\plotone{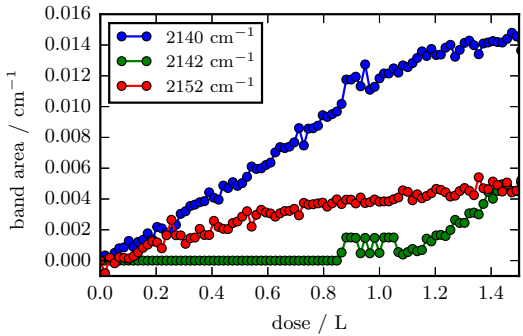}
\caption{Band area versus deposition dose for the absorption peaks in Figure~\ref{fig:co_com_spectrum}.\label{fig:co_com_area}}
\end{figure}

Similar experiments were carried out for CO on np-ASW\@. \citet{Lauck2015} have already found that the diffusion of CO becomes significant at $>15$~K. We kept the np-ASW at 22~K, at which the diffusion is fast, while the desorption is still negligible \citep{He2016b}. Based on previous measurement from our group \citep{He2016a}, at 22~K the sticking of CO is unity. CO was deposited continuously from the beam at a rate of 0.035 L$\cdot$minute$^{-1}$. The measured IR spectra are shown in Figure~\ref{fig:co_com_spectrum}. Three different absorption bands can be seen in the figure, at 2140 \cm{}, 2142 \cm{}, and 2152 \cm{}.
The assignment of these peaks are discussed extensively in \citet{Palumbo2006a,Palumbo2006b,Cuppen2011}. At low coverage there are only two peaks at 2140 \cm{} and 2152 \cm{}, growing linearly with coverage. The former has been seen towards numerous young stellar objects \citep{Pontoppidan2003a} and is due to the interaction of CO with non-dangling bond sites. The latter is due to CO interacting with dangling-bond (dOH) sites \citep{Al-Halabi2004}. After $\sim$0.6 L, the band at 2152 \cm{} saturates while the 2140 \cm{} peak continues to increase linearly until about 1.1$\pm0.1$ L, at which the peak at 2142 \cm{} due to pure CO (CO in CO environment) emerges. We used two Gaussian distributions to fit the components at 2152 and 2140 \cm{}, and one Lorentzian distribution to fit the component at 2142 \cm{} \citep{He2016c}. The corresponding areas of these three components are shown in Figure~\ref{fig:co_com_area}.

For CO the critical coverage at which CO in the pure form emerges is at $1.1\pm0.1$ L, which is much higher than that for \cd{}. Later it will be shown that 1~L corresponds to 1 ML\@. This suggests that CO completely wets the surface before building up layers. The underlying reason is because the CO-CO interaction is weaker than the CO-water interaction. On the contrary, the \cd{}-\cd{} interaction is stronger than the \cd{}-water interaction and \cd{} forms clusters. This is a well-known case in thin film growth, also called Vollmer-Weber growth. On the other extreme of the adsorbate-adsorbate versus adsorbate-substrate interaction, growth occurs layer-by-layer. This is the so called Frank-van der Merwe growth \citep[e.g.][]{Ratsch2003}.

In surface science literature, one of the tools to study the formation of islands or clusters on a surface is TPD\@. TPD spectra are usually analyzed using the Polanyi-Wigner equation:
\begin{equation}
\frac{\dif \theta(t)}{\dif t}= -\nu \theta^n
(t) \exp \left(-\frac{E_{\mathrm{b}} (\theta)}{k_{\mathrm{B}}T(t)}\right)
\label{eq:p-w}
\end{equation}
where $\nu$ is the desorption pre-exponential factor (or prefactor). In this work we take the widely used value of $10^{12}$~s$^{-1}$.  $\theta (t)$ is the coverage defined as percentage of  a monolayer (ML), i.e., the number of adsorbate particles divided by the number of adsorption sites on the surface, $n$ is the order of desorption, $E_{\mathrm{b}}(\theta)$ is the binding energy, which can depend on coverage, $k_{\mathrm{B}}$ is Boltzmann constant, $T(t)$ is the temperature of the surface. For $n=0$, the desorption rate does not depend on the amount of material on the surface. Thus, this indicates desorption from multilayer films. $n=0$ or close to it applies also in the desorption of clusters of particles on the surface. Zeroth order desorption is recognized by the fact that TPD traces of film with different values of the coverage have overlapping leading edges \citep{Kolasinski2008}.

\begin{figure}
\plotone{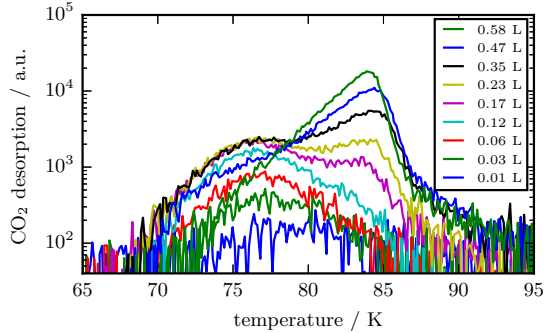}
\caption{Temperature programmed desorption (TPD) spectra of $^{13}$CO$_2$ on np-ASW\@. The deposition dose is shown in the inset. $^{13}$CO$_2$ was deposited from the beam when the surface was at 50~K. The heating ramp rate during the TPD is 0.5~K/s.\label{fig:CO2_TPD}}
\end{figure}

\begin{figure}
\plotone{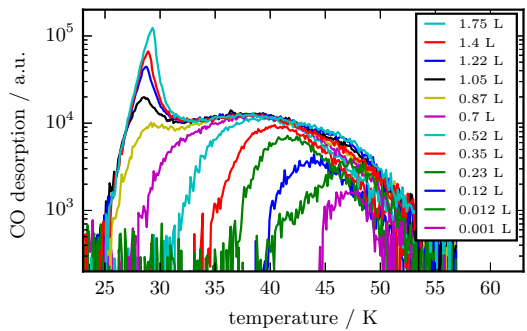}
\caption{Same as in Figure~\ref{fig:CO2_TPD} but for CO on np-ASW\@.  CO was deposited from the beam when the surface was at 22~K.\label{fig:CO_TPD}}
\end{figure}

We carried out TPDs of both  \cd{} and CO on np-ASW, see  Figure~\ref{fig:CO2_TPD} and \ref{fig:CO_TPD}, respectively. For \cd{}, $^{13}$CO$_2$ was used to obtain higher signal-to-noise ratio. From here on the isotope label is dropped. In the TPD experiments for both CO and \cd{}, the deposition rate is 0.70 L$\cdot$minute$^{-1}$, and the heating ramp rate is 0.5~K$\cdot$s$^{-1}$. These TPD spectra of \cd{} and CO are similar to our previous measurement \citep{He2016b}, except here the signal-to-noise ratio is better. The \cd{} TPDs here are for submonolayer coverage; for higher and multilayer coverage, see \citet{He2016b}.
In the case of CO on np-ASW at a coverage  from zero to 0.7~L, the TPD traces show a similar shape and the peak temperature shifts to lower values with increasing deposition. This translates to a decrease in the binding energy with coverage. Between 0.7 L and 0.87 L, a second peak emerges, but it does not have the characteristics of zeroth order desorption yet, such as a common leading edge. At this point, the interaction of  CO is with both H$_2$O and CO\@. Above 1.05~L, the peak at $\sim$ 30 K is due to the CO-CO interaction, as is evident from the common leading edge, indicating multilayer growth. In the submonolayer regime, trailing edges overlap, which indicates that CO occupies deep adsorption sites before shallow sites. We applied the same direct inversion procedure as in \citet{He2016b} to obtain the desorption energy distribution (not shown); we find that it is  close to the distribution reported in \citet{He2016b}. The binding energy of CO-CO (870~K) is located at or below the lower boundary of CO-H$_2$O binding energy (870--1600~K). At about 1.05~L zeroth order desorption is observed, which suggests that a multilayer CO is building up. This coincides with the infrared measurements shown in Figure~\ref{fig:co_com_spectrum}. We define 1~ML coverage as the amount of CO molecules that fully covers the surface area on np-ASW, and in the case  of CO on np-ASW\@ 1~ML corresponds to $4.2\times10^{14}$molecules$\cdot$cm$^{-2}$ (see Section~\ref{sec:exp}). This number may also be used for other molecules after taking into account the correction factor for the size of molecules.

In the \cd{} TPDs, from 0.01~L to 0.06~L there is only one peak centered at $\sim$76~K with small variation with coverage, which indicates first order desorption. At 0.12~L, a second peak at $\sim$ 84~K emerges and continues to grow as coverage increases. This second peak is due to the formation of \cd{} clusters.
At coverage higher than 0.35~L, the first peak at 76~K begins to drop (see traces at 0.47~L and  0.58~L). This is because at higher coverage, as \cd{} cluster formation begins, there is an decreasing number of isolated \cd{} molecules and an increasing number of \cd{} molecules in clusters. At even higher coverages, the TPD shows features of typical zeroth order desorption \citep{He2016b}.

\begin{figure}
\plotone{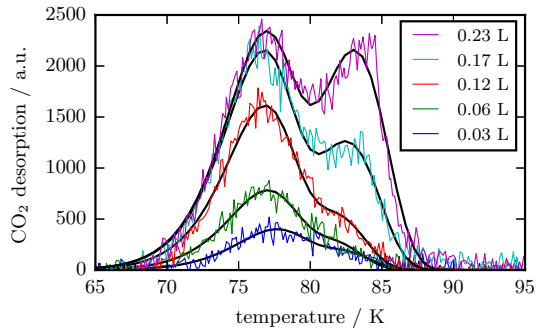}
\caption{Fitting of selected TPD spectra in Figure~\ref{fig:CO2_TPD} using two first order desorption peaks. The experimental data are shown in thin color lines and fitted curves in thick black lines.\label{fig:CO2_TPD_fit}}
\end{figure}

\begin{figure}
\plotone{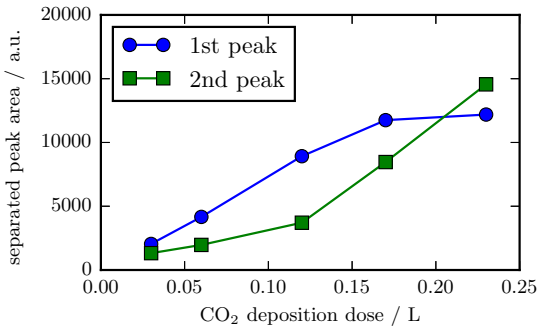}
\caption{The area of the two TPD peaks obtained from the fitting shown in Figure~\ref{fig:CO2_TPD_fit}. Lines are only to guide the eye. \label{fig:CO2_TPD_area}}
\end{figure}

To obtain \cd{}-H$_2$O and \cd{}-\cd{} interaction energies, we fitted the low coverage \cd{} TPDs in Figure~\ref{fig:CO2_TPD} using two first order desorption peaks. Strictly speaking, the second peak due to CO$_2$ desorption from CO$_2$ clusters is not first order. But when the coverage is much lower than the coverage at which zero order desorption is significant, it is a good approximation. We only fit TPDs up to 0.23~L because at higher coverages the fitting is no longer satisfactory. The fitting result is shown in Figure~\ref{fig:CO2_TPD_fit}. The best fitting energies are $2250\pm20$~K and $2415\pm20$~K for the first peak (\cd{}-H$_2$O) and second peak (\cd{}-\cd{}), respectively.
This \cd{}-H$_2$O interaction energy value is comparable with the value of 2268~K on the same type of ice obtained by \citet{Noble2012}.  Figure~\ref{fig:CO2_TPD_area} shows the area of the two peaks with coverage. The first peak for discrete \cd{} shows saturation after $\sim$0.17~L. The second peak for clusters grows slowly at very low coverage and becomes faster when the first peak saturates.

\section{Discussion and Astrophysical Implications}\label{sec:dastro}
Following the observation of pure \cd{} in low mass young stellar objects \citep{Pontoppidan2008}, it has been proposed \citep{Gerakines1999,Pontoppidan2008} that \cd{} could be used as a tracer of the temperature of the ice. \citet{Pontoppidan2008,Ioppolo2009,Ioppolo2013} showed there are two ways to explain the aggregation of \cd{} in ices; it is either by segregation---which involves diffusion in the ice matrix, or by distillation, that is the sublimation of CO in CO-\cd{} mixed layers. Using the data of \citet{Ehrenfreund1999}, they came up with an activation energy for segregation of 4900~K. In new IR studies of \cd{}-water ice mixtures,  \citet{Oberg2009} fitted the amount of segregated \cd{} as a function of time using an Arrhenius expression,  finding an activation energy of 1080$\pm$190~K and a pre-exponent of $2\times10^{5\pm1}$ for a H$_2$O:\cd{}=2:1 mixture. The very high concentration of \cd{} prevents them to single out the effect of surface diffusion.
Our investigation is instead centered on cluster formation at much lower concentration of \cd{} in water ice. Therefore,  it complements the work of \citet{Oberg2009}.

To extract the diffusion energy barrier $E_{\rm dif}$ for a single \cd{} molecule on the surface of np-ASW, we developed a simple rate equation model to simulate the formation of clusters. We assume that $E_{\rm dif}$ takes a single value instead of a continuous distribution. We assume that clusters do not diffuse. The rate equation deals only with diffusion and ignores desorption. Consequently, we only attemp to fit $T<70$~K part of the experiment.  The diffusion rate is expressed as:
\begin{equation}
  D(t)=\nu\exp(-E_{\rm dif}/k_{\rm B}T(t))
\end{equation}
where $\nu$ is the pre-exponential factor for diffusion, which is assumed to be the same as desorption prefactor $10^{12}$~s$^{-1}$. This is a reasonable assumption considering that we describe the motion of a single small molecule largely unaffected by the presence of other \cd{} molecules. We denote $C_1(t)$ as the coverage of isolated \cd{} molecules at time $t$; this is also the surface density of isolated \cd{} molecules divided by the density of adsorption sites. Similarly, the coverage of \cd{} clusters with size  $i$ is denoted by $C_i(t)$. The total density of \cd{} in clusters divided by the density of adsorption sites is $\sum_{i>1} iC_i(t)$. In the simulation we consider $i$ up to 4.  We therefore have the following rate equations:
\begin{align}
\frac{\dif C_1(t)}{\dif t} & =-(2C_1(t)+C_2(t)+C_3(t))C_1(t)D(t)\\
\frac{\dif C_2(t)}{\dif t} & =(C_1(t)-C_2(t))C_1(t)D(t)\\
\frac{\dif C_3(t)}{\dif t} & =(C_2(t)-C_3(t))C_1(t)D(t)\\
\frac{\dif C_4(t)}{\dif t} & =C_3(t)C_1(t)D(t)
\end{align}

\begin{figure}
\plotone{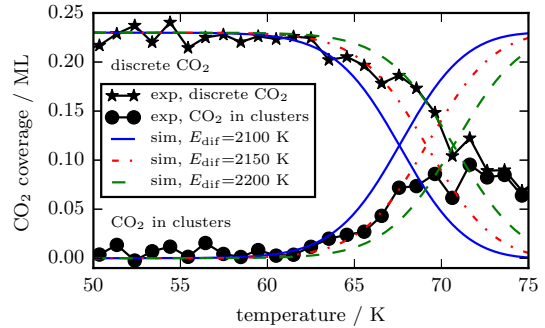}
\epsscale{1.1}
\caption{Comparison of measured and simulated amount of \cd{} in clusters and as isolated molecules on np-ASW\@ expressed in units of fractions of a monolayer. The measured amount of \cd{} is obtained from the integration of the two peaks in Figure~\ref{fig:co2_com_6}; it is then normalized to the \cd{} coverage. The simulation was performed for three different values of the diffusion energy barrier: $E_{\rm dif}=2100, 2150, 2200$~K.\label{fig:diff_sim}}
\end{figure}

Based on this simple rate equation model, we run simulations for the experiment shown in Figure~\ref{fig:co2_com_6}. The initial condition of the simulation is taken to be  $C_1(0)=0.23$ and $C_{i>1}(0)=0$, where 0.23~ML is the initial surface coverage. Figure~\ref{fig:diff_sim} shows the comparison between experimental and simulation results of the coverage of \cd{} both in clusters and as isolated molecules. The measured coverages are normalized to the coverage 0.23~ML\@. At above 70~K, \cd{} desorption begins, which is not simulated by the model. It can been seen that $E_{\rm dif}=2150\pm50$~K fits the experimental data well.

We performed a calculation similar to that in \citet{Oberg2009} to predict the segregation temperature $T_{\rm seg}$ at an astrophysical relevant timescale. We assumed three different heating rates of dust grains, $10^{-2}$, $10^{-3}$, and $10^{-4}$~K per year, which covers a wider range than in a typical warm-up model. The energy barrier for diffusion is $2150\pm50$~K. We assume that on average it takes each \cd{} molecule 5 diffusion steps to encounter another \cd{} and segregate or form clusters, which is a reasonable estimate considering the \cd{} abundance with respect to water. It is found that the segregation temperature $T_{\rm seg}=43\pm3$~K. This is higher than the value obtained by \citet{Oberg2009} who assumed a heating rate of 30~K in 4,000 years and a \cd{} concentration of  0.16 in water ice.

In summary, we found that the diffusion energy barrier for a \cd{} molecule diffusing on np-ASW is 2150$\pm$50~K, assuming a diffusion prefactor of $10^{12}$ s$^{-1}$. A calculation shows that at astrophysical timescales the \cd{} segregation temperature is $43\pm3$~K. This temperature is higher than the one at which there is CO distillation in CO:\cd{} ice \citep{Pontoppidan2008}. Assuming the same prefactor for desorption, we also obtained the binding energies of \cd{} on \cd{} ice and \cd{} on np-ASW to be $2415\pm20$~K and $2250\pm20$~K, respectively. A stronger binding of \cd{}-\cd{} than \cd{}-H$_2$O is consistent with the non-wetting of \cd{} on water ice. These values can then be used in simulations of the formation and evolution of mixed ices in the ISM\@.

\section{Acknowledgments}
This research was supported by NSF Astronomy \& Astrophysics Division Grants \#1311958 and \#1615897.

\section{Appendix}
\begin{figure}
\plotone{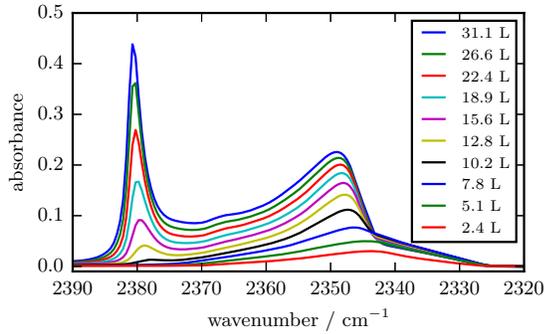}
\caption{CO$_2$ $\nu_3$ band during background deposition of \cd{} on 100~L of water ice that is deposited on gold surface at 10~K and warmed  to 67 K.}
\label{fig:calibration}
\end{figure}

The deposition of \cd{} on np-ASW is obtained either from filling the chamber at a given pressure for a given time (background deposition) or using the molecular beam. The molecular beam deposition is slow and has a narrow angular spread, thus it is targeted to the sample with minimal deposition of \cd{} molecules on other parts of the apparatus. The deposition rate from background $R_{bg}$ can be calculated by
  \begin{eqnarray}
    R_{bg}=nv/4\\
    n=P/k_{\rm B}T
  \end{eqnarray}
where $n$ is the density in the gas phase, $v$ is the velocity of the gas particle, $P$ is the pressure in the chamber, and $T$ is the gas temperature.  The gas pressure was assumed to be homogeneous in the chamber. The gas specific correction factor of ionization gauge was already taken into account. The temperature of the gas was taken to be the same as the chamber wall.
The deposition rate from the beam was obtained by comparing the IR of \cd{} deposited from the beam with that from background deposition. In the calibration experiments, \cd{} was deposited on 100~L of water that was deposited at 10~K and warmed to 67~K.  At 67~K \cd{} is mobile on the water surface. RAIRS spectra were taken during \cd{} deposition as shown in the Figure~\ref{fig:calibration}. At increasing coverage, the 2344 cm$^{-1}$ feature moves to higher wavenumber, while the 2380 cm$^{-1}$ peak begins to emerge (for clarity, only  traces at selected exposures are shown). The benchmark coverage was chosen to be the one at which the 2344 \cm{} peak is at the same height as the 2380 \cm{} peak. This benchmark coverage is independent of deposition method---background and beam deposition. It only depends on the coverage of \cd{} on the part of the water ice that is covered by \cd{}.  Therefore this is a reliable method to obtain the absolute intensity of the beam. The relative uncertainty of the beam intensity calibration is 5\%.

\begin{thebibliography}{}
\expandafter\ifx\csname natexlab\endcsname\relax\def\natexlab#1{#1}\fi

\bibitem[{{Al-Halabi} {et~al.}(2004){Al-Halabi}, {Fraser}, {Kroes}, \& {van
  Dishoeck}}]{Al-Halabi2004}
{Al-Halabi}, A., {Fraser}, H.~J., {Kroes}, G.~J., \& {van Dishoeck}, E.~F.
  2004, \aap, 422, 777

\bibitem[{{Baratta} {et~al.}(2000){Baratta}, {Palumbo}, \&
  {Strazzulla}}]{Baratta2000}
{Baratta}, G.~A., {Palumbo}, M.~E., \& {Strazzulla}, G. 2000, \aap, 357, 1045

\bibitem[{{Bernstein} {et~al.}(2005){Bernstein}, {Cruikshank}, \&
  {Sandford}}]{Bernstein2005}
{Bernstein}, M.~P., {Cruikshank}, D.~P., \& {Sandford}, S.~A. 2005, Icar, 179,
  527

\bibitem[{{Berreman}(1963)}]{Berreman1963}
{Berreman}, D.~W. 1963, PhRv, 130, 2193

\bibitem[{{Boogert} {et~al.}(2015){Boogert}, {Gerakines}, \&
  {Whittet}}]{Boogert2015}
{Boogert}, A.~C.~A., {Gerakines}, P.~A., \& {Whittet}, D.~C.~B. 2015, \araa,
  53, 541

\bibitem[{{Cook} {et~al.}(2011){Cook}, {Whittet}, {Shenoy}, {Gerakines},
  {White}, \& {Chiar}}]{Cook2011}
{Cook}, A.~M., {Whittet}, D.~C.~B., {Shenoy}, S.~S., {et~al.} 2011, \apj, 730,
  124

\bibitem[{{Cuppen} {et~al.}(2011){Cuppen}, {Penteado}, {Isokoski}, {van der
  Marel}, \& {Linnartz}}]{Cuppen2011}
{Cuppen}, H.~M., {Penteado}, E.~M., {Isokoski}, K., {van der Marel}, N., \&
  {Linnartz}, H. 2011, \mnras, 417, 2809

\bibitem[{{Dartois} {et~al.}(1999){Dartois}, {Schutte}, {Geballe}, {Demyk},
  {Ehrenfreund}, \& {D'Hendecourt}}]{Dartois1999}
{Dartois}, E., {Schutte}, W., {Geballe}, T.~R., {et~al.} 1999, \aap, 342, L32

\bibitem[{{Edridge} {et~al.}(2013){Edridge}, {Freimann}, {Burke}, \&
  {Brown}}]{Edridge2013}
{Edridge}, J.~L., {Freimann}, K., {Burke}, D.~J., \& {Brown}, W.~A. 2013,
  RSPTA, 371, 20110578

\bibitem[{{Ehrenfreund} {et~al.}(1998){Ehrenfreund}, {Dartois}, {Demyk}, \&
  {D'Hendecourt}}]{Ehrenfreund1998}
{Ehrenfreund}, P., {Dartois}, E., {Demyk}, K., \& {D'Hendecourt}, L. 1998,
  \aap, 339, L17

\bibitem[{{Ehrenfreund} {et~al.}(1999){Ehrenfreund}, {Kerkhof}, {Schutte},
  {Boogert}, {Gerakines}, {Dartois}, {D'Hendecourt}, {Tielens}, {van Dishoeck},
  \& {Whittet}}]{Ehrenfreund1999}
{Ehrenfreund}, P., {Kerkhof}, O., {Schutte}, W.~A., {et~al.} 1999, \aap, 350,
  240

\bibitem[{{Escribano} {et~al.}(2013){Escribano}, {Munoz Caro}, {Cruz-Diaz},
  {Rodriguez-Lazcano}, \& {Mate}}]{Escribano2013}
{Escribano}, R.~M., {Munoz Caro}, G.~M., {Cruz-Diaz}, G.~A.,
  {Rodriguez-Lazcano}, Y., \& {Mate}, B. 2013, PNAS, 110, 12899

\bibitem[{{Gerakines} \& {Hudson}(2015)}]{Gerakines2015}
{Gerakines}, P.~A., \& {Hudson}, R.~L. 2015, \apjl, 808, L40

\bibitem[{{Gerakines} {et~al.}(1999){Gerakines}, {Whittet}, {Ehrenfreund},
  {Boogert}, {Tielens}, {Schutte}, {Chiar}, {van Dishoeck}, {Prusti},
  {Helmich}, \& {de Graauw}}]{Gerakines1999}
{Gerakines}, P.~A., {Whittet}, D.~C.~B., {Ehrenfreund}, P., {et~al.} 1999,
  \apj, 522, 357

\bibitem[{{Ghesqui{\`e}re} {et~al.}(2015){Ghesqui{\`e}re}, {Mineva}, {Talbi},
  {Theul{\'e}}, {Noble}, \& {Chiavassa}}]{Ghesquiere2015}
{Ghesqui{\`e}re}, P., {Mineva}, T., {Talbi}, D., {et~al.} 2015, PCCP, 17, 11455

\bibitem[{{He} {et~al.}(2016{\natexlab{a}}){He}, {Acharyya}, \&
  {Vidali}}]{He2016b}
{He}, J., {Acharyya}, K., \& {Vidali}, G. 2016{\natexlab{a}}, \apj, 825, 89

\bibitem[{{He} {et~al.}(2016{\natexlab{b}}){He}, {Acharyya}, \&
  {Vidali}}]{He2016a}
---. 2016{\natexlab{b}}, \apj, 823, 56

\bibitem[{He {et~al.}(2016)He, Emtiaz, \& Vidali}]{He2016c}
He, J., Emtiaz, S., \& Vidali, G. 2016, in prep.

\bibitem[{{Hodyss} {et~al.}(2008){Hodyss}, {Johnson}, {Orzechowska}, {Goguen},
  \& {Kanik}}]{Hodyss2008}
{Hodyss}, R., {Johnson}, P.~V., {Orzechowska}, G.~E., {Goguen}, J.~D., \&
  {Kanik}, I. 2008, Icar, 194, 836

\bibitem[{{Ioppolo} {et~al.}(2009){Ioppolo}, {Palumbo}, {Baratta}, \&
  {Mennella}}]{Ioppolo2009}
{Ioppolo}, S., {Palumbo}, M.~E., {Baratta}, G.~A., \& {Mennella}, V. 2009,
  \aap, 493, 1017

\bibitem[{{Ioppolo} {et~al.}(2013){Ioppolo}, {Sangiorgio}, {Baratta}, \&
  {Palumbo}}]{Ioppolo2013}
{Ioppolo}, S., {Sangiorgio}, I., {Baratta}, G.~A., \& {Palumbo}, M.~E. 2013,
  \aap, 554, A34

\bibitem[{{Isokoski} {et~al.}(2014){Isokoski}, {Bossa}, {Triemstra}, \&
  {Linnartz}}]{Isokoski2014}
{Isokoski}, K., {Bossa}, J.-B., {Triemstra}, T., \& {Linnartz}, H. 2014, PCCP,
  16, 3456

\bibitem[{{Karssemeijer} {et~al.}(2014){Karssemeijer}, {de Wijs}, \&
  {Cuppen}}]{Karssemeijer2014}
{Karssemeijer}, L.~J., {de Wijs}, G.~A., \& {Cuppen}, H.~M. 2014, PCCP, 16,
  15630

\bibitem[{{Keane} {et~al.}(2001){Keane}, {Boogert}, {Tielens}, {Ehrenfreund},
  \& {Schutte}}]{Keane2001}
{Keane}, J.~V., {Boogert}, A.~C.~A., {Tielens}, A.~G.~G.~M., {Ehrenfreund}, P.,
  \& {Schutte}, W.~A. 2001, \aap, 375, L43

\bibitem[{Kolasinski(2008)}]{Kolasinski2008}
Kolasinski, K. 2008, Surface Science: Foundations of Catalysis and Nanoscience
  (Wiley)

\bibitem[{{Lauck} {et~al.}(2015){Lauck}, {Karssemeijer}, {Shulenberger},
  {Rajappan}, {{\"O}berg}, \& {Cuppen}}]{Lauck2015}
{Lauck}, T., {Karssemeijer}, L., {Shulenberger}, K., {et~al.} 2015, \apj, 801,
  118

\bibitem[{{Mispelaer} {et~al.}(2013){Mispelaer}, {Theul{\'e}}, {Aouididi},
  {Noble}, {Duvernay}, {Danger}, {Roubin}, {Morata}, {Hasegawa}, \&
  {Chiavassa}}]{Mispelaer2013}
{Mispelaer}, F., {Theul{\'e}}, P., {Aouididi}, H., {et~al.} 2013, \aap, 555,
  A13

\bibitem[{{Noble} {et~al.}(2012){Noble}, {Congiu}, {Dulieu}, \&
  {Fraser}}]{Noble2012}
{Noble}, J.~A., {Congiu}, E., {Dulieu}, F., \& {Fraser}, H.~J. 2012, \mnras,
  421, 768

\bibitem[{{Nummelin} {et~al.}(2001){Nummelin}, {Whittet}, {Gibb}, {Gerakines},
  \& {Chiar}}]{Nummelin2001}
{Nummelin}, A., {Whittet}, D.~C.~B., {Gibb}, E.~L., {Gerakines}, P.~A., \&
  {Chiar}, J.~E. 2001, \apj, 558, 185

\bibitem[{{{\"O}berg} {et~al.}(2009){{\"O}berg}, {Fayolle}, {Cuppen}, {van
  Dishoeck}, \& {Linnartz}}]{Oberg2009}
{{\"O}berg}, K.~I., {Fayolle}, E.~C., {Cuppen}, H.~M., {van Dishoeck}, E.~F.,
  \& {Linnartz}, H. 2009, \aap, 505, 183

\bibitem[{{Palumbo}(2006)}]{Palumbo2006b}
{Palumbo}, M.~E. 2006, \aap, 453, 903

\bibitem[{{Palumbo} \& {Baratta}(2000)}]{Palumbo2000}
{Palumbo}, M.~E., \& {Baratta}, G.~A. 2000, \aap, 361, 298

\bibitem[{{Palumbo} {et~al.}(2006){Palumbo}, {Baratta}, {Collings}, \&
  {McCoustra}}]{Palumbo2006a}
{Palumbo}, M.~E., {Baratta}, G.~A., {Collings}, M.~P., \& {McCoustra}, M.~R.~S.
  2006, Physical Chemistry Chemical Physics (Incorporating Faraday
  Transactions), 8, 279

\bibitem[{{Pontoppidan} {et~al.}(2003){Pontoppidan}, {Fraser}, {Dartois},
  {Thi}, {van Dishoeck}, {Boogert}, {d'Hendecourt}, {Tielens}, \&
  {Bisschop}}]{Pontoppidan2003a}
{Pontoppidan}, K.~M., {Fraser}, H.~J., {Dartois}, E., {et~al.} 2003, \aap, 408,
  981

\bibitem[{{Pontoppidan} {et~al.}(2008){Pontoppidan}, {Boogert}, {Fraser}, {van
  Dishoeck}, {Blake}, {Lahuis}, {{\"O}berg}, {Evans}, \&
  {Salyk}}]{Pontoppidan2008}
{Pontoppidan}, K.~M., {Boogert}, A.~C.~A., {Fraser}, H.~J., {et~al.} 2008,
  \apj, 678, 1005

\bibitem[{{Ratsch} \& {Venables}(2003)}]{Ratsch2003}
{Ratsch}, C., \& {Venables}, J.~A. 2003, JVST, 21, S96

\bibitem[{{Whittet} {et~al.}(2009){Whittet}, {Cook}, {Chiar}, {Pendleton},
  {Shenoy}, \& {Gerakines}}]{Whittet2009}
{Whittet}, D.~C.~B., {Cook}, A.~M., {Chiar}, J.~E., {et~al.} 2009, \apj, 695,
  94

\bibitem[{{Whittet} {et~al.}(2007){Whittet}, {Shenoy}, {Bergin}, {Chiar},
  {Gerakines}, {Gibb}, {Melnick}, \& {Neufeld}}]{Whittet2007}
{Whittet}, D.~C.~B., {Shenoy}, S.~S., {Bergin}, E.~A., {et~al.} 2007, \apj,
  655, 332

\bibitem[{{Whittet} {et~al.}(1998){Whittet}, {Gerakines}, {Tielens}, {Adamson},
  {Boogert}, {Chiar}, {de Graauw}, {Ehrenfreund}, {Prusti}, {Schutte},
  {Vandenbussche}, \& {van Dishoeck}}]{Whittet1998}
{Whittet}, D.~C.~B., {Gerakines}, P.~A., {Tielens}, A.~G.~G.~M., {et~al.} 1998,
  \apjl, 498, L159

\end{thebibliography}

\end{document}